\documentclass[12pt,twoside,a4paper]{article}
\pdfoutput=1

\usepackage{amsmath, amsthm, amsfonts, amssymb}
\usepackage{graphicx}
\usepackage{float}
\usepackage{color}
\usepackage{cite}
\usepackage[width=\textwidth,font=small,labelfont=bf,
labelsep=endash]{caption}
\usepackage{titlesec}

\titleformat{\section}
  {\normalfont\fontsize{13}{16}\bfseries}{\thesection}{1em}{}
\titleformat{\subsection}
  {\normalfont\fontsize{11}{14}\bfseries}{\thesubsection}{1em}{}

\normalsize
\newcommand\Lame {Lam\'e\ }
\newcommand\Backlund {B\"{a}cklund }
\newcommand\Schrodinger {Schr\"{o}dinger }

\begin{document}

\title{\textbf{Stability Analysis of Classical String Solutions and the Dressing Method}}
\author{Dimitrios Katsinis$^{1,2}$, Ioannis Mitsoulas$^3$ and Georgios Pastras$^2$}
\date{\small $^1$Department of Physics, National and Kapodistrian University of Athens,\\University Campus, Zografou, Athens 15784, Greece\\
$^2$NCSR ``Demokritos'', Institute of Nuclear and Particle Physics,\\Aghia Paraskevi 15310, Attiki, Greece\\
$^3$Department of Physics, School of Applied Mathematics and Physical Sciences,\\National Technical University, Athens 15780, Greece\linebreak \vspace{8pt}
\texttt{dkatsinis@phys.uoa.gr, mitsoula@central.ntua.gr, pastras@inp.demokritos.gr}}

\vskip .5cm

\maketitle

\begin{abstract}
The dressing method is a technique to construct new solutions in non-linear sigma models under the provision of a seed solution. This is analogous to the use of auto\Backlund transformations for systems of the sine-Gordon type. In a recent work, this method was applied in the sigma model that describes string propagation on $\mathbb{R} \times \mathrm{S}^2$, using as seeds the elliptic classical string solutions. Some of the new solutions that emerge reveal instabilities of their elliptic precursors \cite{salient}. The focus of the present work is the fruitful use of the dressing method in the study of the stability of closed string solutions. It establishes an equivalence between the dressing method and the conventional linear stability analysis. More importantly, this equivalence holds true in the presence of appropriate periodicity conditions that closed strings must obey. Our investigations point to the direction of the dressing method being a general tool for the study of the stability of classical string configurations in the diverse class of symmetric spacetimes.
\newline \newline \textbf{Keywords:} Classical Strings, Stability of Classical Strings, Dressing Method, Pohlmeyer Reduction
\end{abstract}

\newpage

\tableofcontents

\newpage

\setcounter{equation}{0}
\section{Introduction}
The dressing method is a tool that can be used in order to generate new non-trivial solutions of non-linear sigma models (NLSMs) using a given seed solution. In the context of NLSMs that describe string propagation on symmetric spaces, the dressing method has been applied extensively \cite{Spradlin:2006wk,Kalousios:2006xy}. The NLSMs on symmetric spaces exhibit an important additional property, due to their integrability, the so called Pohlmeyer reduction. Pohlmeyer reduction turns the Gauss-Codazzi equations for the embedding of the world-sheet in the target space, which is in turn embedded in an enhanced flat space, to an integrable system of the family of the sine-Gordon equation. Many properties of the solutions of the NLSM can be understood much easier in the context of the corresponding Pohlmeyer reduced theory. 

Based on the inversion of the Pohlmeyer reduction, we have recently constructed classical string solutions on $\mathbb{R}\times \mathrm{S}^2$ \cite{part1}. Our approach resulted in the classification of the elliptic solutions of this particular NLSM. The convenient parametrization of this approach, facilitates the application of the dressing method to the whole class of these solutions. The physics of the dressed solutions is very rich, since many new effects arise \cite{salient}, such as spike interactions. Most of these effects could not, and obviously have not been, found in the preceding literature, due to the fact that the dressing method had been applied either to solutions that correspond to the vacuum of the Pohlmeyer reduced theory, or to soliton solutions.

One interesting finding of \cite{salient} is the existence of a particular class of solutions, which correspond to instabilities of closed elliptic strings. The periodicity conditions that are obeyed by the closed strings, combined with the physics of the dressed strings, specify a particular region of parameters in the moduli space of the elliptic string solutions that allow the existence of these instabilities. This work deals with the relation of this class of dressed solutions to linearized perturbations around the elliptic ones. We establish a one-to-one correspondence between the instabilities of the linearized perturbations around the Pohlmeyer field of the seed solution and the dressed solutions that realize the instabilities of the elliptic strings. As a consequence, the dressing method can be a useful tool for the study of string instabilities.

The structure of the paper is as follows: In section \ref{sec:review}, we review some basic facts, firstly on the construction of the elliptic string solutions via the inversion of the Pohlmeyer reduction, and secondly on the dressed elliptic string solutions, which realize instabilities of their elliptic seeds. In section \ref{sec:linear}, we study the linear perturbations of the elliptic strings in the language of the Pohlmeyer reduced system and show that there is a one-to-one correspondence of unstable linear perturbations and the relevant dressed string solutions. In section \ref{sec:moduli}, we specify explicitly the set of unstable elliptic string solutions and finally, in section \ref{sec:discussion}, we discuss our results.

\setcounter{equation}{0}
\section{Dressed Elliptic Stings}
\label{sec:review}

Many interesting classical string solutions on $\mathbb{R} \times \mathrm{S}^2$ have been found and studied. Some well-known examples include the GKP string \cite{GKP_string}, the BMN particle \cite{BMN}, the giant magnon \cite{Giant_Magnons} and the single spike \cite{single_spike_rs2}. A more extended class of solutions are the helical (spiky) strings \cite{dual_spikes,helical,multi,Kruczenski:2006pk}, which can be shown to include the above simpler solutions as special limits \cite{part1}.

A convenient method to construct and study these solutions is through the Pohlmeyer reduction \cite{Pohlmeyer:1975nb,Zakharov:1973pp}. The Pohlmeyer reduction is a non-trivial, non-local mapping between NLSMs defined on symmetric target spaces to integrable theories of the family of the sine-Gordon equation, the so called Symmetric Space Sine-Gordon models (SSSGs). The NLSM, which describes string propagation on $\mathbb{R} \times \mathrm{S}^2$, is Pohlmeyer reducible to the sine-Gordon equation itself. This mapping between the two theories provides a simple way to classify the classical string solutions: the helical strings correspond to elliptic solutions of the sine-Gordon equation, i.e. they are trains of kinks or kink-antikink pairs. These solutions have the additional property that they uniquely depend on one of the two worldsheet coordinates at some appropriate frame.

The use of the sine-Gordon counterparts allows the generic study of the elliptic string solutions on $\mathbb{R} \times \mathrm{S}^2$ through the inversion of the Pohlmeyer reduction. This procedure, which is highly non-trivial in general due to the non-local nature of Pohlmeyer reduction, can be performed in the case of elliptic solutions. It results in simple expressions for the latter in terms of the Weierstrass elliptic function $\wp$, which facilitate the study of their properties \cite{part1}, namely,
\begin{align}
t_{0/1} &= \sqrt {{x_2} - \wp \left( a \right)} {\xi ^0} + \sqrt {{x_3} - \wp \left( a \right)} {\xi ^1} , \label{eq:elliptic_solutions_t} \\
\vec X_{0/1} &= \left( {\begin{array}{*{20}{c}}
{{F_1}\left( {{\xi^{0/1}}} \right)\cos \left(\ell {\xi^{1/0}} - \Phi \left( \xi^{0/1} ; a \right)\right)}\\
{{F_1}\left( {{\xi^{0/1}}} \right)\sin \left(\ell {\xi^{1/0}} - \Phi \left( \xi^{0/1} ; a \right)\right)}\\
{{F_2}\left( {{\xi^{0/1}}} \right)}
\end{array}} \right) ,
\label{eq:elliptic_solutions_review}
\end{align}
where
\begin{equation}
{F_1}\left( {{\xi}} \right) = \sqrt {\frac{{\wp \left( {{\xi} + {\omega _2}} \right) - \wp \left( a \right)}}{{{x_1} - \wp \left( a \right)}}} , \quad {F_2}\left( {{\xi}} \right) = \sqrt {\frac{{{x_1} - \wp \left( {{\xi} + {\omega _2}} \right)}}{{{x_1} - \wp \left( a \right)}}} , \label{eq:dressed_strings_F1_F2} 
\end{equation}
and
\begin{equation}
{\Phi }\left( {{\xi };a} \right) :=  - \frac{i}{2}\ln \frac{{\sigma \left( {{\xi } + {\omega _2} + a} \right)\sigma \left( {{\omega _2} - a} \right)}}{{\sigma \left( {{\xi } + {\omega _2} - a} \right)\sigma \left( {{\omega _2} + a} \right)}} + i\zeta \left( a \right){\xi } .
\label{eq:elliptic_solutions_lame_phase_def}
\end{equation}
The functions $\zeta$ and $\sigma$ are the Weierstrass quasi-periodic functions. The moduli of the Weierstrass elliptic and associated functions are given by
\begin{equation}
g_2 = \frac{E^2}{3} + \mu^4 , \quad g_3 = \frac{E}{3} \left( \left( \frac{E}{3} \right)^2 - \mu^4 \right) .
\label{eq:elliptic_solutions_moduli}
\end{equation}
The quantities $x_1$, $x_2$ and $x_3$ are the three roots of the cubic polynomial associated to the Weierstrass elliptic function. They are always real and they are given by
\begin{equation}
x_1 = \frac{E}{3} , \quad x_{2/3} = - \frac{E}{6} \pm \frac{\mu^2}{2}.
\end{equation}
In the following, we will use a twofold notation for these roots. The symbols $x_i$ denote the roots as defined by the above equations. Additionally, we will use the notation $e_i$ for the ordered roots, i.e. $e_1>e_2>e_3$. It holds true that whenever $-\mu^2<E<\mu^2$, then $x_2>x_1>x_3$, whereas whenever $E>\mu^2$, we have $x_1>x_2>x_3$; the parameter $E$ cannot take values smaller than $-\mu^2$.

These solutions form a two-parameter family, which is parametrized by $E$ and $a$. The Pohlmeyer reduction is a many-to-one mapping; only the first parameter affects the Pohlmeyer counterpart, i.e. solutions that have the same $E$ and different $a$ have the same Pohlmeyer counterpart (modulo a worldsheet boost).

It is well-known that given one solution of the sine-Gordon equation, one can construct new ones, via the application of a \Backlund transformation. This is a system of two first order differential equations, which connects pairs of solutions. The dressing method constitutes the equivalent to this procedure in the framework of NLSMs \cite{Zakharov:1980ty,Harnad:1983we,Pohl_avatars}. In general, it is highly non-trivial to apply the dressing method for the construction of new solutions. In the past, it has been applied only for the case of the seed being very simple (the BMN particle) \cite{Spradlin:2006wk,Kalousios:2006xy}; it gets mapped to the vacuum of the sine-Gordon equation via the Pohlmeyer reduction. However, the simplified expressions for the elliptic strings obtained in \cite{part1} allow for the construction of even more complicated solutions, via the application of the dressing method \cite{Katsinis:2018ewd}.

In general, the application of a single \Backlund transformation increases the genus of the solution by one. This extra genus is degenerate, i.e. one of the related periods in the complex moduli space diverges. This naturally means that the action of the \Backlund transformation corresponds to the insertion of a kink on the background of the seed solution. Indeed, the most well studied example of generation of new solutions via the application of a \Backlund transformation is the construction of the single kink solutions, using the vacuum as the seed. In the case of an elliptic seed, the situation is similar \cite{salient}.

In \cite{salient}, it was shown that depending on a specific property of the seed solution, it is possible that the divergent period introduced by the \Backlund transformation on the sine-Gordon side can be aligned with the temporal direction on the NLSM side. Moreover, the new solution inherits the periodicity properties of the seed and hence, as long as the seed is a well-defined closed string solution, the corresponding new solution becomes one as well. In such cases, the new dressed solutions have a very interesting property: Asymptotically in time, they are identical to the elliptic seed (or actually a rotated version of the seed), whereas in intermediate times they are completely different. These solutions reveal that the seed elliptic string is unstable. The dressed solution is the analogue of the trajectory that connects two unstable equilibrium positions in a system such as the simple pendulum.

The condition for the existence of such solutions is the following: The elliptic string has a Pohlmeyer counterpart that is elliptic too. This means that in a specific worldsheet frame the solution depends solely on one of the two worldsheet coordinates. However, this frame is not necessarily the static gauge i.e. the timelike worldsheet coordinate does not coincide with the physical time. This is evident in equation \eqref{eq:elliptic_solutions_t}. Actually, the special case where it does, corresponds to the GKP limit. This implies that there is a boost with a specific given velocity $\beta < 1$,
\begin{equation}
\beta = \sqrt{\frac{x_3 - \wp \left( a \right)}{x_2 - \wp \left( a \right)}} ,
\label{eq:boost_velocity}
\end{equation}
which has to be performed in order to write the solution in the static gauge. On the other hand, the dressed solution, as we analysed above, has a Pohlmeyer counterpart which is a kink propagating on top on the elliptic background. This kink, in the frame where the background depends on a single worldsheet coordinate travels with a specific given mean velocity ${\bar v} _{0/1}$, 
\begin{equation}
{\bar v} _0 = \frac{{\zeta \left( {\tilde a} \right){\omega _1} - \zeta \left( {{\omega _1}} \right)\tilde a}}{{{\omega _1}D}} , \quad
{\bar v} _1 = \frac{{{\omega _1}D}}{{\zeta \left( {\tilde a} \right){\omega _1} - \zeta \left( {{\omega _1}} \right)\tilde a}}.
\label{eq:kinks_mean_velocity}
\end{equation}
The indices $0$ and $1$ denote that the elliptic seed depends solely on the time-like or space-like worldsheet coordinate in an appropriate frame respectively. In other words the index $0$ is used in the case of translationally invariant seed and the index $1$ in the case of static seeds. The parameters $D$ and $\tilde{a}$ are determined by the location of the poles of the dressing factor or equivalent by the \Backlund parameter. They are connected as
\begin{equation}
D^2 = \wp \left( \tilde{a} \right) - x_1 .
\label{eq:D_atilde}
\end{equation}
More details are provided in \cite{Katsinis:2018ewd,salient}. This mean velocity may be smaller or larger than the speed of light. The condition of existence of a dressed elliptic string, which reveals an instability of the seed elliptic solution is that its sine-Gordon elliptic counterpart can host a superluminal kink with mean velocity equal to
\begin{equation}
{\bar v} _{0/1} = - \frac{1}{\beta} .
\label{eq:instability_condition}
\end{equation}

The elliptic strings can be classified into four classes in terms of the generic properties of their Pohlmeyer counterparts. These are characterized by whether the Pohlmeyer counterpart is static or translationally invariant in an appropriate worldsheet frame, and furthermore by whether they are oscillatory (a train of kink-antikink pairs) or rotating (a train of kinks). At the level of the moduli, the oscillatory solutions are characterized by $E<\mu^2$, whereas the rotating solutions by $E>\mu^2$. At the level of the corresponding roots, oscillatory and rotating solutions are distinguished by the ordering of the roots $x_i$. It turns out that the instabilities can be found in all static oscillatory elliptic solutions and in no static rotating ones, whereas some translationally invariant solutions exhibit the instabilities, too.

It follows that the stability of a classical string solution can be studied via the use of the dressing method. However, the dressing method reveals some specific instabilities; it is not necessary that the string solutions without these instabilities are actually stable. In order to better understand the relation between the dressing method and the actual stability of the string solution, we have to compare the above results to a standard linear stability analysis. This is the subject of the next section.

\setcounter{equation}{0}
\section{Perturbations of the Pohlmeyer Field}
\label{sec:linear}

As we discussed in the previous section, the NLSM describing the propagation of strings on $\mathbb{R} \times \mathrm{S}^2$ is Pohlmeyer reducible to the sine-Gordon equation. This turns out to be handy whenever one wants to study the stability of a classical string solution. Instead of studying the stability of the original NLSM solution, one may study the stability of its Pohlmeyer counterpart, which is much simpler.

The stability of the elliptic solutions of the sine-Gordon equation has been studied in the literature \cite{SGstability}. It turns out that all elliptic solutions are unstable except for the static rotating ones. However, this analysis has been performed without taking into account the necessary periodic conditions that a solution of the sine-Gordon equation must obey in order to correspond to a closed string on the NLSM side. We will perform this analysis in the following.

The sine-Gordon equation reads
\begin{equation}
\partial_1^2\varphi(\xi^0,\xi^1)-\partial_0^2\varphi(\xi^0,\xi^1)=\mu^2\sin\varphi(\xi^0,\xi^1) .
\end{equation}
Assume we are given a solution of the sine-Gordon equation $\bar{\varphi}(\xi^0,\xi^1)$. We introduce a perturbed solution of the form
\begin{equation}
\varphi(\xi^0,\xi^1)=\bar{\varphi}(\xi^0,\xi^1)+\tilde{\varphi}(\xi^0,\xi^1),
\end{equation}
where $\tilde{\varphi}(\xi^0,\xi^1) \ll \bar{\varphi}_0(\xi^0,\xi^1)$. For notational convenience we drop the arguments of the fields in what follows. At linear order the perturbations obey the following equation
\begin{equation}\label{eq:pohlmeyer_elliptic}
\partial_1^2\tilde{\varphi}-\partial_0^2\tilde{\varphi}=\mu^2\left(\cos\bar{\varphi}\right)\tilde{\varphi} .
\end{equation}
A general remark is that this equation has exactly the same form as the equation obeyed by the embedding functions of the string solution. We restrict our attention to the case where $\bar{\varphi}$ is an elliptic solution. Such solutions have the property that they depend on a sole world-sheet coordinate in an appropriate frame. We abbreviate $\bar{\varphi}(\xi^{0/1})=\bar{\varphi}_{0/1}$ and denote as $\tilde{\varphi}_{0/1}$ the corresponding perturbation. For perturbations around elliptic solutions, one can implement separation of variables and subsequently turn the equation into a pair of effective \Schrodinger problems. One of the two problems is trivial, whereas the other one corresponds to the integrable $n=1$ \Lame potential. For details see \cite{part1}.

The elliptic solutions in the frame where they depend on only one worldsheet coordinate assume the form
\begin{equation}
\cos\bar{\varphi}_{0/1}=\mp\frac{1}{\mu^2}\left(2\wp\left(\xi^{0/1}+\omega_2\right)+x_1\right) .
\end{equation}
Equation \eqref{eq:pohlmeyer_elliptic} turns to
\begin{equation}
\partial_1^2\tilde{\varphi}_{0/1}-\partial_0^2\tilde{\varphi}_{0/1}=\mp\left(2\wp\left(\xi^{0/1}+\omega_2\right)+x_1\right)\tilde{\varphi}_{0/1} .
\end{equation}
This equation can be solved via separation of variables. It has solutions of the form
\begin{align}
\tilde{\varphi}_0=e^{i k \xi^{1}}y_0(\xi_0),\\
\tilde{\varphi}_1=e^{i \omega \xi^{0}}y_1(\xi_1),
\end{align}
where the functions $y_{0/1}$ satisfy the following equations
\begin{align}
-\partial_0^2y_0+2\wp\left(\xi^0+\omega_2\right)y_0=\left(k^2-x_1\right)y_0,\\
-\partial_1^2y_1+2\wp\left(\xi^1+\omega_2\right)y_1=\left(\omega^2-x_1\right)y_1 ,
\end{align} 
which are the infamous $n=1$ \Lame problem. Our analysis is based on the band structure of the \Lame potential, which can be analytically studied along with its properties.

Before we proceed, two crucial remarks are in order. So far we have used a linear gauge for the physical time, \eqref{eq:elliptic_solutions_t}. This choice facilitates the separation of variables. It is a key element of our approach in constructing classical string solutions in $\mathbb{R}\times \mathrm{S}^2$. Yet, the study of the string should be performed in the static gauge, since it is in this gauge that the physical time coincides with the timelike worldsheet coordinate. The second remark is the relation of the string worldsheet to the domain of the sine-Gordon equation. The boundary conditions of closed strings imply that the Pohlmeyer field should be periodic with respect to the spacelike worldsheet coordinate in the static gauge, which is equivalent to considering the sine-Gordon equation in $\mathbb{R}\times \mathrm{S}^1$. This compactification assigns well defined topological charge to the solutions and sets strict constraints on the linear perturbations.

\subsection{The band structure of the $n=1$ \Lame potential}
We will present some basic facts about the band structure of the \Lame potential. We refer the reader to \cite{bakas_pastras,Pastras:2017wot} for details.

The eigenfunctions of the $n=1$ \Lame equation
\begin{equation}
-\partial^2 y+2\wp\left(\xi+\omega_2\right)y=\lambda y
\end{equation}
are
\begin{equation}
y_\pm(\xi)=\frac{\sigma\left(\xi+\omega_2\pm \tilde{a} \right)\sigma\left(\omega_2\right)}{\sigma\left(\xi+\omega_2\pm\right)\sigma\left(\omega_2\pm \tilde{a} \right)}e^{-\zeta(\pm\tilde{a})\xi},
\end{equation}
where the corresponding eigenvalue is $\lambda=-\wp(\tilde{a})$. The symbol $\tilde{a}$ used in this section should not be confused with the modulus $\tilde{a}$ of the dressed string solutions presented in the previous section. However, we use the same symbol as it will turn out in what follows that they coincide. The eigenvalue $\lambda$ should be taken real, so that the perturbations are also real for any value of both worldsheet coordinates. This constraints $\tilde{a}$ to take values on a specific domain on the complex plane. Since the roots of the cubic polynomial associated to the Weierstrass elliptic function are always real, the Weierstrass elliptic function has a real period $2\omega_1$ and a purely imaginary period $2\omega_2$. Then, the fundamental domain, where the Weierstrass elliptic function is real, is the union of four linear segments (obviously it is real in all other positions that are congruent to this fundamental domain). These are the segments connecting the origin to $\omega_1$ and $\omega_2$, as well as the segments connecting $\omega_3 = \omega_1 + \omega_2$ to the latter. These form a closed rectangle. The Weierstrass elliptic function assumes any real value at exactly one point on this rectangle. More specifically, on the segment connecting the origin to $\omega_1$ it decreases monotonously from $+\infty$ to the largest root as $\tilde{a}$ moves from the latter to the former. Similarly, on the segment connecting $\omega_1$ to $\omega_3$ it assumes any value between the two largest roots. On the segment connecting $\omega_3$ to $\omega_2$ it assumes any value between the two smallest roots. Finally, on the segment connecting $\omega_2$ to the origin it assumes any value smaller than the smallest root.

When $\tilde{a}$ lies in the segment connecting the origin to $\omega_1$ or the segment connecting $\omega_2$ to $\omega_3$ the eigenfunctions $y_\pm$ are both real and unbounded, while in the other two cases, the eigenfunctions $y_\pm$ are complex conjugate to each other and bounded, corresponding to Bloch waves of the $n=1$ \Lame potential.

In all cases, the eigenfunctions obey the quasi-periodicity property
\begin{equation}
y_\pm(\xi+2\omega_1)=y_\pm(\xi)e^{\pm2\left(\zeta(\omega_1)\tilde{a}-\zeta(\tilde{a})\omega_1\right)}.
\end{equation}
This quasiperiodicity property of the eigenfunction is of critical importance. It is determined by the behaviour of the function $f(z)=\zeta(\omega_1)z-\zeta(z)\omega_1$ on the domain where the Weierstrass elliptic function is real. This function has the following properties
\begin{enumerate}
\item
When $z$ lies on the segment defined by $0$ and $\omega_1$, the function f is real. As $z$ moves from $0$ to $\omega_1$, $f(z)$ increases monotonously from $-\infty$ to $0$.
\item
When $z$ lies on the segment defined by $\omega_1$ and $\omega_3$, the function f is purely imaginary. As $z$ moves from $\omega_1$ to $\omega_3$ the imaginary part of $f(z)$ increases monotonously from $0$ to $\pi/2$.
\item
When $z$ lies on the segment defined by $\omega_3$ and $\omega_2$, the function f is complex. Its imaginary part is constant and equals to $\pi/2$; as $z$ moves from $\omega_3$ to $\omega_2$ its real part decreases monotonously from $0$ to a minimum value and then increases monotonously to $0$.
\item
Finally, when $z$ lies on the segment defined by $\omega_2$ and $0$, the function f is purely imaginary. As  $z$ moves from $\omega_2$ to $0$, its imaginary part increases monotonously from $\pi/2$ to $+\infty$.
\end{enumerate}

\subsection{Perturbations of Closed Strings}

The static gauge, which is defined by $t=\mu \sigma^0$, is connected to the linear gauge \eqref{eq:elliptic_solutions_t} via an approapriate boost with velocity given by \eqref{eq:boost_velocity}. In this gauge, the perturbations of static elliptic solutions assume the following form
\begin{equation}\label{eq:perturbations_st}
\tilde{\varphi}_1(\sigma^0,\sigma^1)=e^{i \omega \gamma\left(\sigma^0-\beta \sigma^1\right)}y_1(\gamma\left(\sigma^1-\beta \sigma^0\right)),
\end{equation}
where $\gamma = 1 / \sqrt{1-\beta^2}$. The parameter $\omega$ is related to the eigenvalue of the \Lame equation as
\begin{equation}
\omega^2=x_1-\wp(\tilde{a}).
\label{eq:omega}
\end{equation}
The perturbation $\tilde{\varphi}_1$ obeys the following quasi-periodicity property
\begin{equation}
\tilde{\varphi}_1(\sigma^0,\sigma^1+2n\omega_1/\gamma)=\tilde{\varphi}_1(\sigma^0,\sigma^1)e^{2n\left(-i\beta\omega_1\omega\pm f(\tilde{a})\right)} .
\end{equation}
Assume that this perturbation corresponds to a small perturbation of a closed string solution, which is covered by $\sigma^1 \in \left[0 , 2 n \omega_1 / \gamma \right)$, where $n \in \mathbb{Z}$. The perturbation should be periodic in $\sigma^1$ with period $2 n \omega_1 / \gamma$, thus, the perturbation should obey the periodicity condition
\begin{equation}\label{eq:pert_periodicity_st}
-i\beta\omega_1\omega+f(\tilde{a})=\frac{m}{n}\pi i ,
\end{equation}
where $m,n \in \mathbb{Z}$. 

Similarly, the perturbations of the translational invariant elliptic solutions are of the form
\begin{equation}\label{eq:perturbations_ti}
\tilde{\varphi}_0(\sigma^0,\sigma^1)=e^{i k \gamma\left(\sigma^1-\beta \sigma^0\right)}y_0(\gamma\left(\sigma^0-\beta \sigma^1\right)).
\end{equation}
The parameter $k$ is related to the eigenvalue of the \Lame equation as
\begin{equation}
k^2=x_1-\wp(\tilde{a}).
\label{eq:kappa}
\end{equation}
The perturbation $\tilde{\varphi}_0$ has the following quasi-periodicity property
\begin{equation}
\tilde{\varphi}_0(\sigma^0,\sigma^1+(2n\omega_1)/(\beta\gamma))=\tilde{\varphi}_0(\sigma^0,\sigma^1)e^{-2n\left(-i\frac{\omega_1 k}{\beta}\pm f(\tilde{a})\right)}.
\end{equation}
Similarly, the appropriate periodicity condition for perturbations of translationally invariant elliptic solutions of the sine-Gordon equation, which correspond to closed string solutions, are 
\begin{equation}\label{eq:pert_periodicity_ti}
-i\frac{\omega_1 k}{\beta}+f(\tilde{a})=\frac{m}{n}\pi i,
\end{equation}
where $m,n \in \mathbb{Z}$.

\subsection{The Time Evolution of the Perturbations}
In the previous section, we derived the conditions that are obeyed by perturbations of closed elliptic strings, namely equations \eqref{eq:pert_periodicity_st} and \eqref{eq:pert_periodicity_ti}. These conditions determine the spectrum of the perturbations defining the admissible values of $\tilde{a}$. Nevertheless the existence of such perturbations is not sufficient in order to analyse the stability properties. One should proceed and study the time evolution of these perturbations. 

The quasiperiodic properties of the solutions determine the time evolution of the perturbations. More specifically, it holds that
\begin{align}
{{\tilde \varphi }_1}\left( {{\sigma ^0} + \frac{{2n{\omega _1}}}{{\gamma \beta }},{\sigma ^1}} \right) &= {{\tilde \varphi }_1}\left( {{\sigma ^0},{\sigma ^1}} \right){e^{n \Delta {\Phi _1}}} , \label{eq:time_quasiperiodicity_st}\\
{{\tilde \varphi }_0}\left( {{\sigma ^0} + \frac{{2n{\omega _1}}}{\gamma },{\sigma ^1}} \right) &= {{\tilde \varphi }_0}\left( {{\sigma ^0},{\sigma ^1}} \right){e^{n \Delta {\Phi _0}}} , \label{eq:time_quasiperiodicity_ti}
\end{align}
where $n \in \mathbb{Z}$ and
\begin{align}
\Delta {\Phi _1} &= 2\left( {i\frac{{\omega {\omega _1}}}{\beta } \mp f \left( {\tilde a} \right)} \right) , \\
\Delta {\Phi _0} &= 2\left( { - i\beta k {\omega _1} \pm f \left( {\tilde a} \right)} \right) .
\end{align}

Whenever the spatial periodicity conditions \eqref{eq:pert_periodicity_st} or \eqref{eq:pert_periodicity_ti} are obeyed, the quasiperiodicity properties \eqref{eq:time_quasiperiodicity_st} and \eqref{eq:time_quasiperiodicity_ti} are obeyed with
\begin{align}
\Delta {\Phi _1} &= 2i \frac{{\omega {\omega _1}}}{{\beta {\gamma ^2}}} ,\\
\Delta {\Phi _0} &= 2i \frac{{k{\omega _1}}}{{\beta {\gamma ^2}}} .
\end{align}
These equations clearly imply that whenever $\omega$ or $k$ are real, the perturbation has an oscillatory and bounded evolution in time, with period
\begin{align}
T _1 &= \frac{{ 2 \pi{\gamma }}}{{\left| \omega \right| }} ,\\
T _0 &= \frac{{2 \pi \beta {\gamma }}}{{\left| k \right|}} .
\end{align}
On the other hand, when $\omega$ or $k$ is imaginary, the perturbations grow exponentially, revealing that the elliptic solution is unstable. In these cases the corresponding Lyapunov exponents are
\begin{align}
\lambda _1 &= \frac{{\left| \omega \right| }}{{ {\gamma }}} ,\\
\lambda _0 &= \frac{{\left| k \right|}}{{\beta {\gamma }}} .
\end{align}

\subsection{Analysis of the Spectrum of the Perturbations}

The parametrization in terms of Weierstrass elliptic functions turns out to be a great advantage, since the whole presentation can be held very short, without relying on a tentative case by case analysis. Table \ref{tab:values_tab} summarizes the range of the function $f(z)=\zeta(\omega_1)z-\zeta(\tilde{a})z$ that is related on the quasi-periodicity of the eigenfunctions of the \Lame equation.
\begin{table}[h]
\centering
\begin{tabular}{|c |c |c| c| c|}
\hline
$\tilde{a}$ & $\wp(\tilde{a})$ & $x_1-\wp(\tilde{a})$ & $f\left( \tilde{a} \right)$ & $c$ \\
\hline\hline
segment defined by $0$ and $\omega_1$ & $\in(e_1,\infty)$ & $-$ & $c$ & $c\in(-\infty,0)$\\
\hline
segment defined by $\omega_1$ and $\omega_3$ & $\in(e_2,e_1)$ & $-/+$ & $i c$ & $c\in(0,\pi/2)$\\
\hline
segment defined by $\omega_3$ and $\omega_2$ & $\in(e_3,e_2)$ & $+$ & $c+i \pi/2$ & $c\in(M,0)$\\
\hline
segment defined by $\omega_2$ and $0$ & $\in(-\infty,e_3)$ & $+$ & $ic$ & $c\in(\pi/2,\infty)$\\
\hline
\end{tabular}
\caption{The values of the parameters, entering the equations \eqref{eq:pert_periodicity_st} and\eqref{eq:pert_periodicity_ti}, for various values of $\tilde{a}$. In the 3rd column, 2nd row the sign $-$ corresponds to oscillating solutions, while the sign $+$ corresponds to rotating solutions.}
\label{tab:values_tab}
\end{table}
\begin{enumerate}
\item When $\tilde{a}$ lies on the segment defined by $0$ and $\omega_1$, $\wp \left( \tilde{a} \right)$ is larger than any of the three roots, thus, equations \eqref{eq:omega} and \eqref{eq:kappa} imply that the parameter $\omega$ or $k$ is imaginary. The left-hand-side of \eqref{eq:pert_periodicity_st} and \eqref{eq:pert_periodicity_ti} is real, as a result only the $m=0$ sector could provide solutions with appropriate periodicity conditions. The spatial periodicity conditions \eqref{eq:pert_periodicity_st} and \eqref{eq:pert_periodicity_ti}, for $m=0$ are equivalent to the condition 
\begin{equation}
\beta = \frac{f \left( \tilde{a} \right)}{i \omega \omega_1} ,
\label{eq:instability_condition_st}
\end{equation}
in the case of a static elliptic solution and
\begin{equation}
\beta = \frac{i k \omega_1}{f \left( \tilde{a} \right)} ,
\label{eq:instability_condition_ti}
\end{equation}
for translationally invariant elliptic solutions.

Let us compare the above to the findings from the dressed elliptic string solution. Taking into account the equations \eqref{eq:kinks_mean_velocity}, \eqref{eq:D_atilde} and the fact that the parameters $\omega$ and $k$ are given by the expressions \eqref{eq:omega} and \eqref{eq:kappa}, the above conditions \eqref{eq:instability_condition_st} and \eqref{eq:instability_condition_ti} become identical to the conditions that emerged from the dressed elliptic solutions \eqref{eq:instability_condition}. This identifies the $\tilde{a}$ parameter of our linear analysis to the $\tilde{a}$ modulus of the kinks that propagate on top of the elliptic background as presented in \cite{Katsinis:2018ewd}.

Whenever such solution can be found, since $\omega$ or $k$ is imaginary, the perturbations grow exponentially in time and reveal that the elliptic solution is unstable.

\item When $\tilde{a}$ lies on the segment defined by $\omega_1$ and $\omega_3$, $\wp \left( \tilde{a} \right)$ lies between the two largest roots. In the case that the elliptic solution is oscillatory ($x_2>x_1>x_3$), following equations \eqref{eq:omega} and \eqref{eq:kappa}, the parameter $\omega$ or $k$ is imaginary and the left-hand-side of equations \eqref{eq:pert_periodicity_st} and \eqref{eq:pert_periodicity_ti} is complex, thus providing no solution with appropriate periodicity conditions. On the other hand, in the rotating case ($x_1>x_2>x_3$) the parameter $\omega$ or $k$ is real, and as a result, the equations \eqref{eq:pert_periodicity_st} and \eqref{eq:pert_periodicity_ti} could possess valid solutions.

Since they are characterized by real parameter $\omega$ or $k$, these perturbations are stable. Interestingly enough, the spacial periodicity condition in this case assumes the same form as the condition of existence of closed dressed elliptic solutions with $D^2<0$, as presented in \cite{salient}. These indeed exist only when the seed is a rotating elliptic solution, similarly to the outcome our this linear analysis.

\item When $\tilde{a}$ lies on the segment defined by $\omega_3$ and $\omega_2$, $\wp \left( \tilde{a} \right)$ lies between the two smallest roots, and, thus, the parameter $\omega$ or $k$ is real. The left-hand-side of \eqref{eq:pert_periodicity_st} and \eqref{eq:pert_periodicity_ti} is complex, as a result they do not possess solutions with appropriate periodicity conditions.

\item Finally, when $\tilde{a}$ lies on the segment defined by $\omega_2$ and $0$, $\wp \left( \tilde{a} \right)$ is smaller than the smallest root and the parameter $\omega$ or $k$ is real similarly to the previous case. The left-hand-side of \eqref{eq:pert_periodicity_st} and \eqref{eq:pert_periodicity_ti} is imaginary. As a result, they could provide valid solutions, in which case the perturbations are also stable, since either $\omega$ or $k$ is real. The spatial periodicity condition assumes a form similar to the appropriate periodicity conditions for the elliptic strings themselves \cite{part1}.
\end{enumerate}

Summing up, the elliptic solutions are unstable, whenever one can find a perturbation with $\tilde{a}$ in the segment connecting the origin and $\omega_1$ that obeys the condition \eqref{eq:instability_condition_st} or \eqref{eq:instability_condition_ti}, for static and translationally invariant elliptic solutions, respectively. The results of our linear analysis are identical to the full non-linear construction of the unstable trajectories with the use of the dressing method \cite{Katsinis:2018ewd,salient}. This strongly supports the dressing method as a tool for the stability analysis of classical string solutions.

\section{The Moduli Space of Unstable Elliptic Solutions}
\label{sec:moduli}

The main purpose of our work, i.e. the demonstration of equivalence between a linear stability analysis and the non-linear construction of unstable trajectories with the use of the dressing method, has been realized. In this section, we will determine the subset of unstable elliptic string solutions in their moduli space parametrized by the constants $E$ and $a$, as introduced in \cite{part1}.

In all cases, the condition for the existence of an instability has been expressed as
\begin{equation}
\beta = - \frac{1}{\bar{v}_{0/1}},
\end{equation}
where $\bar{v}_{0/1}$ is given by \eqref{eq:kinks_mean_velocity}. Thus, we must study the dependence of $\bar{v}_{0/1}$ on $\tilde{a}$ in order to specify whether there are unstable perturbations of the elliptic strings.

The mean kink velocity $\bar{v}_0$ on a translational invariant background is
\begin{equation}
\bar{v}_0=\frac{\zeta(\tilde{a})\omega_1-\zeta(\omega_1)\tilde{a}}{\omega_1 D}.
\label{eq:kink_velocity_ti_s4}
\end{equation}
Depending on the values of $\tilde{a}$ and $E$, this velocity can be either superluminal or subluminal. We are going to prove that this velocity is strictly superluminal in the case of solutions with a rotating counterpart, whereas there exists a critical value $E_c$ for the moduli $E$, obeying $0<E_c<\mu^2$, such that the kinks on an oscillating background are strictly subluminal for $E<E_c$.

Without loss of generality, let us consider the case $ 0 \leq \tilde{a}\leq \omega_1$. It is a simple task to determine the limits of the velocity as $\tilde{a}$ tends to the endpoints of its allowed region. It holds true that
\begin{equation}
\mathop {\lim }\limits_{\tilde a \to 0} {\bar v}_0 = 1 .
\label{eq:v_at_0}
\end{equation}
In the case of an oscillating background, one can show that
\begin{equation}
\mathop {\lim }\limits_{\tilde a \to \omega_1} {\bar v}_0 = 0 .
\label{eq:v_at_omega1}
\end{equation}
In the case of rotating backgrounds though, the expression for the velocity \eqref{eq:kink_velocity_ti_s4} is undetermined at the limit $\tilde a \to \omega_1$. Expanding appropriately the numerator and the denominator yields
\begin{equation}
\mathop {\lim }\limits_{\tilde a \to {\omega _1}} {\bar v}_0 = \frac{{{{\zeta \left( {{\omega _1}} \right)}} / {{{\omega _1}}} + {x_1}}}{{\sqrt {\left( {{x_1} - {x_2}} \right)\left( {{x_1} - {x_3}} \right)} }} \equiv \bar v_{\max} > 1 .
\label{eq:elliptic_kink_vmax}
\end{equation}
In the vicinity of $\tilde{a}\rightarrow 0^+$ it holds
\begin{equation}\label{eq:expansion_v}
\bar{v}_0=1+c_2 \left(E \right) \tilde{a}^2+\mathcal{O}\left(\tilde{a}^4\right),
\end{equation}
where
\begin{equation}
c_2 \left(E \right) = \frac{x_1}{2}-\frac{\zeta(\omega_1)}{\omega_1} .
\end{equation}

The addition formula of the Weierstrass $\zeta$ function implies that
\begin{equation}
2\zeta(\omega_1)=\zeta\left(z+2\omega_1\right)-\zeta\left(z\right)=-\int_{z}^{z+2\omega_1}\wp(x)dx.
\end{equation}
The Weierstrass elliptic function assumes real values on the segment connecting $\omega_2$ and $\omega_3$ whose range is between the two smallest roots $e_2$ and $e_3$. Thus, for $z=\omega_2$, we obtain
\begin{equation}
\frac{\zeta(\omega_1)}{\omega_1}=-\frac{1}{2\omega_1}\int_0^{2\omega_1}\wp(x+\omega_2)dx.
\end{equation}
We recall that the Weierstrass elliptic function on the segment connecting $\omega_2$ and $\omega_3$ ranges between the two smallest roots. Therefore, it is evident that
\begin{equation}\label{eq:zeta_bound}
-e_2\leq \frac{\zeta(\omega_1)}{\omega_1} \leq -e_3.
\end{equation}
In the case of an oscillating background ($e_2 = x_1$), the above relation can be reexpressed as
\begin{equation}
\frac{3 x_1}{2} \geq c_2 \left(E \right) \geq - \frac{\mu^2}{2} ,
\end{equation}
which implies trivially that
\begin{align}
c_2 \left(E \right) < 0, \quad \textrm{when } 0 > E \geq - \mu^2.
\end{align}
Thus, the second order term in the expansion of the kink velocity with $\tilde{a}$ is negative, whenever the constant $E$ is negative. However, it is not possible to derive its sign with such simple arguments when $E>0$. This is the subject of what follows.

Let us specify the extrema of the kink velocity. Its derivative with respect to $\tilde{a}$ is given by
\begin{equation}
\frac{\partial \bar{v}_0}{\partial \tilde{a}}=-\frac{\wp(\tilde{a})+\frac{\zeta(\omega_1)}{\omega_1}}{\sqrt{\wp(\tilde{a})-x_1}}-\frac{\wp^\prime(\tilde{a})\left(\zeta(\tilde{a})-\frac{\zeta(\omega_1)}{\omega_1}\tilde{a}\right)}{2\left(\wp(\tilde{a})-x_1\right)^{3/2}}.
\label{eq:mean_vel_derivative}
\end{equation}
The absence of a linear term in \eqref{eq:expansion_v} obviously implies that
\begin{equation}
\frac{\partial \bar{v}_0}{\partial \tilde{a}}\Bigg\vert_{\tilde{a}=0}=0.
\end{equation}
At the other endpoint of the possible values of $\tilde{a}$, the derivative of the mean kink velocity depends on whether the elliptic solution is oscillatory or rotating. In both cases, the Weierstrass elliptic function assumes the value of the largest root at $\tilde{a} = \omega_1$. In the case of the solution being oscillatory, $x_1$ is not the largest root, and one can directly read the derivative from \eqref{eq:mean_vel_derivative}. If the solution is rotating, the expression \eqref{eq:mean_vel_derivative} will become indeterminate at $\omega_1$. An appropriate expansion of this formula around $\omega_1$ shows that the derivative vanishes in this case. Thus,
\begin{equation}
\frac{\partial \bar{v}_0}{\partial \tilde{a}}\Bigg\vert_{\tilde{a}=
\omega_1}=\begin{cases}
0, & E>\mu^2 , \\
-\frac{x_2+\frac{\zeta(\omega_1)}{\omega_1}}{\sqrt{x_2-x_1}}, & \mu^2>E>-\mu^2 .
\end{cases}
\end{equation}

In order to study other possible extrema points of the kink velocity, we reexpress its derivative as
\begin{equation}
\frac{\partial \bar{v}_0}{\partial \tilde{a}}=\frac{g(\tilde{a})\wp^\prime(\tilde{a})}{2\left(\wp(\tilde{a})-x_1\right)^{3/2}},
\end{equation}
where
\begin{equation}
g(\tilde{a})=-\frac{2}{\wp^\prime(\tilde{a})}\left(\wp(\tilde{a})+\frac{\zeta(\omega_1)}{\omega_1}\right)\left(\wp(\tilde{a})-x_1\right)-\left(\zeta(\tilde{a})-\frac{\zeta(\omega_1)}{\omega_1}\tilde{a}\right).
\end{equation}
The technical advantage of expressing the derivative of the velocity in this form, is that $g^\prime$  is an elliptic function of $\tilde{a}$ and in particular it is a function of $\wp(\tilde{a})$. Further zeros of the derivative of $\bar{v}_0$ for $\tilde{a}\in(0,\omega_1)$ are solutions of the equation $g(\tilde{a})=0.$ The expansions of the Weierstrass elliptic functions at $\tilde{a}=0$ imply that
\begin{equation}
g(0)=0,
\end{equation}
while for $\tilde{a}=\omega_1$ it holds that
\begin{equation}
g(\omega_1)= \begin{cases}
0, & E>\mu^2, \\
+\infty, & \mu^2>E>-\mu^2 .
\end{cases}
\end{equation}
We are going to study the monotonicity of $g$ in order to specify the number of the solutions of the equation $g(\tilde{a})=0$. It is a matter of algebra to show that the equation $g^\prime(\tilde{a})=0$ is a quadratic equation for $\wp(\tilde{a})$, with solutions
\begin{align}
\wp(\tilde{a})&=x_1 , \label{eq:solfpa1}\\
\wp(\tilde{a})&=\frac{x_2x_3-\frac{x_1}{2}\frac{\zeta(\omega_1)}{\omega_1}}{\frac{\zeta(\omega_1)}{\omega_1}-\frac{x_1}{2}} = - \frac{x_2x_3-\frac{x_1}{2}\frac{\zeta(\omega_1)}{\omega_1}}{c_2 \left(E \right)} . \label{eq:solfpa2}
\end{align}

For $E>\mu^2$, the root $x_1$ is the largest root, thus the first solution \eqref{eq:solfpa1} corresponds trivially to $\tilde{a} = \omega_1$, which is not interesting, since we are looking for solutions in the open interval $( 0 , \omega_1 )$. The second solution \eqref{eq:solfpa2} assumes the form
\begin{equation}\label{eq:wpa_rot}
\wp(\tilde{a})=e_1+\frac{-\frac{3x_1}{2}c_2 \left(E \right)+\frac{\mu^4}{4}}{c_2 \left(E \right)}.
\end{equation}
When $E>\mu^2$ the function $g(\tilde{a})$ is smooth for every $\tilde{a}\in[0,\omega_1]$ and furthermore one can show that $g(0)=g(\omega_1)=0$. Rolle's theorem states that the equation $g^\prime(\tilde{a})=0$ must have at least one solution for $\tilde{a}\in(0,\omega_1)$. Our analysis suggests that \eqref{eq:wpa_rot} is the only possible solution. As a result it is the sole solution whenever $E>\mu^2$. The function $\wp(\tilde{a})$ is real and larger than $e_1$ for every $\tilde{a}\in(0,\omega_1)$. The quantity $c_2 \left(E \right)$ cannot vanish, since $\wp(\tilde{a})$ is finite in $(0,\omega_1)$. If it were negative, the numerator of the fraction in equation \eqref{eq:wpa_rot} would be positive. As a result $\wp(\tilde{a})$ would be smaller than $e_1$. Therefore, we deduce by contradiction that
\begin{equation}\label{eq:coeff_rot}
c_2 \left(E \right) > 0, \quad \textrm{when } E>\mu^2.
\end{equation}

Since $g^\prime$ vanishes only once in the interval $(0,\omega_1)$, it is evident that $g$, and as a consequence $\frac{\partial \bar{v}_0}{\partial \tilde{a}}$, retain their sign. By taking into account \eqref{eq:coeff_rot}, the expansion \eqref{eq:expansion_v} implies that $\frac{\partial \bar{v}_0}{\partial \tilde{a}}>0$, in the region of $\tilde{a} = 0$. Consequently $\bar{v}_0$ is an increasing function of $\tilde{a}$ in the whole interval $(0,\omega_1)$, whenever $E\geq \mu^2$. Taking into account equation \eqref{eq:v_at_0}, this implies that all kinks, which propagate on a translationally invariant rotating elliptic background, possess superluminal velocity.

We are left with the oscillating case, namely $-\mu^2<E<\mu^2$. In this case, the analysis is more complicated. Once again the solution \eqref{eq:solfpa1} is irrelevant since $x_1=e_2$. This would imply that it corresponds to $\tilde{a} = \omega_3$, which does not lie on the segment connecting the origin and $\omega_1$. In the oscillating case, the second solution \eqref{eq:solfpa2} may be reexpressed as
\begin{equation}\label{eq:wpa_osc}
\wp( \tilde{a} )=e_1+\frac{\mu^2}{2}\frac{e_1+\frac{\zeta(\omega_1)}{\omega_1}}{c_2 \left(E \right)}.
\end{equation}
The equation \eqref{eq:zeta_bound} suggests that the numerator of the fraction appearing in \eqref{eq:wpa_osc} is always positive. Hence the solution \eqref{eq:solfpa2} provides a valid $\tilde{a}$ (i.e. $\wp( \tilde{a} )$ is larger than $e_1$), as long as the denominator $c_2 \left(E \right)$ is also positive. We already know that $c_2 \left(E \right)$ is negative whenever $E<0$ and positive whenever $E>\mu^2$. We are going to prove that $c_2 \left(E \right)$ is a monotonous function of $E$ for $-\mu^2<E<\mu^2$. As a result there exists only one critical value of energy $E_c$, such that
\begin{equation}
c_2 \left(E_c \right)=0 .
\end{equation}
This critical value can be numerically found to be equal to $E_c = 0.65223$.

The following formulas are needed:
\begin{align}
\frac{\partial \omega_1}{\partial g_2}=\frac{18 g_3 \zeta(\omega_1)-g_2^2\omega_1}{4\left(g_2^3-27g_3^2\right)}, & \quad \frac{\partial \omega_1}{\partial g_3}=\frac{9 g_3 \omega_1-6g_2 \zeta(\omega_1)}{2\left(g_2^3-27g_3^2\right)},\\
\frac{\partial \zeta(\omega_1)}{\partial g_2}=\frac{2 g_2^2 \zeta(\omega_1)-3g_2g_3\omega_1}{8\left(g_2^3-27g_3^2\right)}, & \quad \frac{\partial \zeta(\omega_1)}{\partial g_3}=\frac{g_2^2 \omega_1-18g_3 \zeta(\omega_1)}{4\left(g_2^3-27g_3^2\right)}.
\end{align}
Using the expressions for the moduli in terms of the constant $E$ \eqref{eq:elliptic_solutions_moduli}, and after some trivial algebra, we obtain:
\begin{align}
\frac{d\omega_1}{dE}&=-\frac{\frac{E}{3}\omega_1+\zeta(\omega_1)}{E^2-\mu^4}\\
\frac{d\zeta(\omega_1)}{dE}&=\frac{\left(E^2+3\mu^4\right)\omega_1+12E\zeta(\omega_1)}{36\left(E^2-\mu^4\right)}.
\end{align}
Since $x_1 = E/3$, it follows that 
\begin{equation}
\frac{d}{dE}\left(\frac{x_1}{2}-\frac{\zeta(\omega_1)}{\omega_1}\right) = \frac{d c_2 \left(E \right)}{dE} = \frac{1}{4} + \frac{\left(\frac{\zeta(\omega_1)}{\omega_1}+\frac{E}{3}\right)^2}{\mu^4 - E^2} >0,\quad \textrm{when } -\mu^2<E<\mu^2 .
\end{equation}

In effect, for $-\mu^2\leq E< E_c$ the equation $g^\prime(\tilde{a})=0$ has no solution, and therefore $g$ and consequently $\frac{\partial \bar{v}_0}{\partial \tilde{a}}$ retain their sign. Since 
\begin{equation}
c_2 \left(E \right) < 0, \quad -\mu^2<E<E_c
\end{equation}
it is clear from the expansion \eqref{eq:expansion_v} that $\bar{v}_0$ is a decreasing function of $\tilde{a}$ in the region of $\tilde{a} = 0$, and, hence it is a decreasing function of $\tilde{a}$ in the whole segment $(0,\omega_1)$, whenever $-\mu^2<E<E_c$. Taking into account the equation \eqref{eq:v_at_0}, we conclude that all kinks, which propagate on a translationally invariant oscillating background, are subluminal for any $\tilde{a}$, whenever $-\mu^2<E<E_c$.

On the contrary, whenever $E_c<E<\mu^2$, there is exactly one extremum of the velocity in $(0, \omega_1 )$ at $\tilde{a} = \tilde{a}_{\max}$, which is given by \eqref{eq:wpa_osc}. The velocity is an increasing function of $\tilde{a}$ in the region of $\tilde{a}=0$, therefore this extremum is a maximum and the corresponding maximum velocity $\bar{v}_0 \left( \tilde{a}_{\max} \right)$ is superluminal. The velocity vanishes at $\tilde{a} = \omega_1$, as follows from equation \eqref{eq:v_at_omega1}. Since it is monotonous in the interval $\left( \tilde{a}_{\max} , \omega_1 \right)$, it follows that it becomes equal to $1$ exactly once, at a critical $\tilde{a}_c$ (which depends on the particular $E$). It follows that, whenever $E_c<E<\mu^2$, the kinks, which propagate on a translationally invariant oscillating background are superluminal for any $0<\tilde{a}<\tilde{a}_c(E)$ and subluminal for any $\tilde{a}_c(E)<\tilde{a}\leq\omega_1$.

As a final comment, in the singular case of $E=\mu^2$, the use of the degenerate form of Weierstrass elliptic functions and some trivial algebra yields
\begin{equation}
\bar{v}_0=\cosh\left(\mu \tilde{a}\right).
\end{equation}

\begin{figure}[ht]
\vspace{10pt}
\begin{center}
\begin{picture}(100,37)
\put(2,1){\includegraphics[width = 0.45\textwidth]{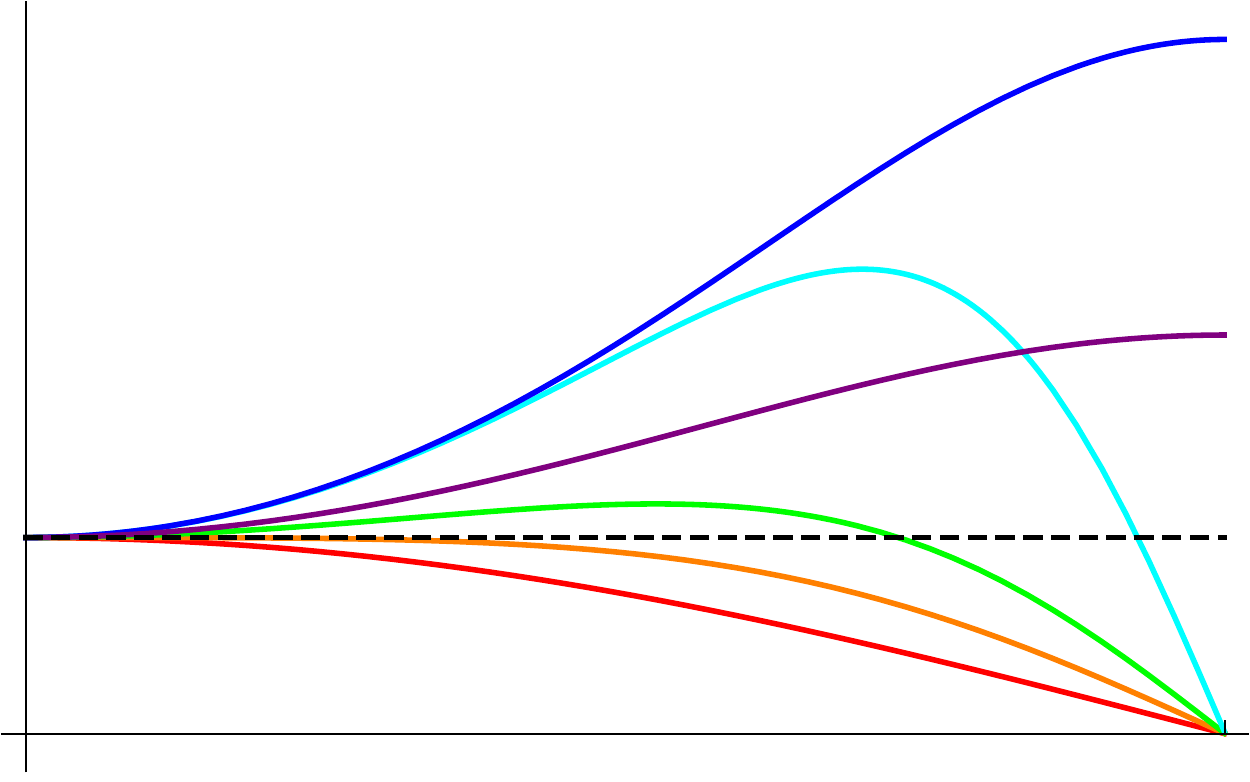}}
\put(45.5,0.5){$1$}
\put(1.5,9){1}
\put(2,30.25){${\bar v}_0$}
\put(47.25,2){$\frac{\tilde{a}}{\omega_1}$}
\put(51,1){\includegraphics[width = 0.45\textwidth]{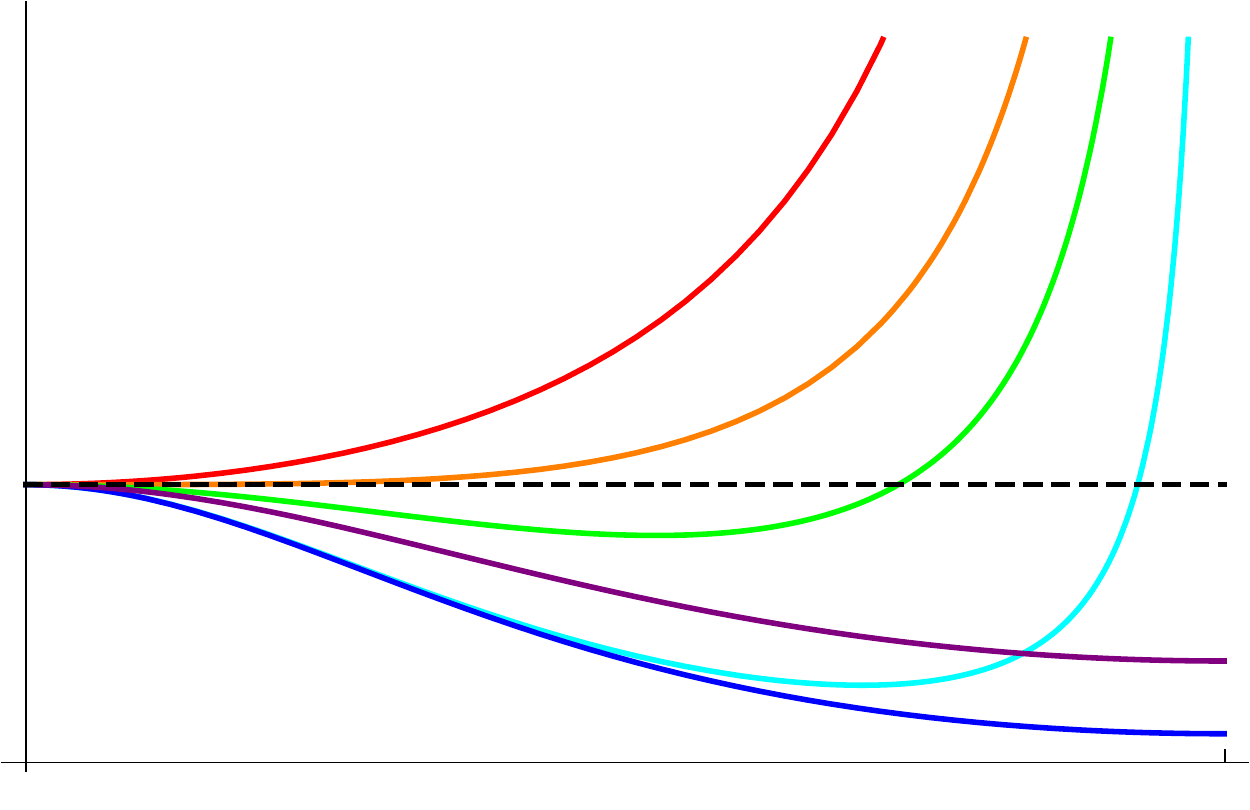}}
\put(94.5,0.5){$1$}
\put(50.5,11.75){1}
\put(51,31){${\bar v}_1$}
\put(96.25,2){$\frac{\tilde{a}}{\omega_1}$}
\put(54,17.75){\includegraphics[height = 0.1875\textwidth]{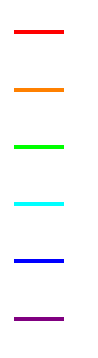}}
\put(54,17.75){\line(0,1){18.75}}
\put(54,17.75){\line(1,0){22.25}}
\put(76.25,17.75){\line(0,1){18.75}}
\put(54,36.5){\line(1,0){22.25}}
\put(58,19){$E=21/20\mu^2$}
\put(58,22){$E=101/100\mu^2$}
\put(58,25){$E=99/100\mu^2$}
\put(58,28){$E=9/10\mu^2$}
\put(58,31){$E=E_c$}
\put(58,34){$E=-9/10\mu^2$}
\end{picture}
\end{center}
\vspace{-10pt}
\caption{The mean velocity as function of $\tilde{a}$ for translationally invariant seeds (left) and static seeds (right) for various values of the energy constant $E$}
\vspace{5pt}
\label{fig:mean_velocity}
\end{figure}
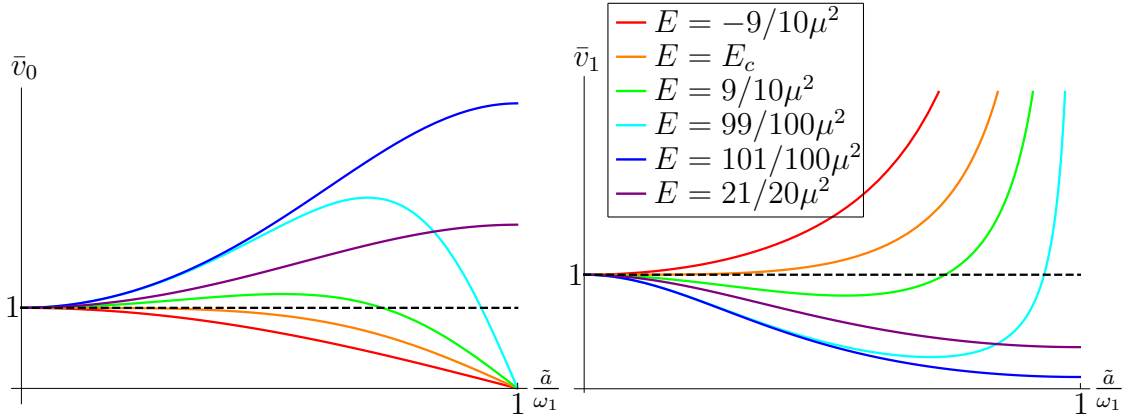

The case of static elliptic solutions can be trivially analyzed, since $\bar{v}_1 = 1 /\bar{v}_0$. Obviously subluminal kinks in the one case correspond to superluminal in the other case and vice versa. The mean kink velocity, as function of $\tilde{a}$ is depicted in figure \ref{fig:mean_velocity}

Returning to the instability of the closed elliptic strings, we recall that these are unstable, whenever it is possible to find a kink propagating on the elliptic background with superluminal mean velocity equal to $1 / \beta$, where $\beta$ is given by \eqref{eq:boost_velocity}. The form of the dependence of the kink mean velocity on $\tilde{a}$ implies the following:
\begin{enumerate}
\item Static oscillatory solutions: The kink mean velocity assumes any value between $1$ and $\infty$ for exactly one value of $\tilde{a}$. Thus, these solutions are \emph{always unstable} and there is \emph{exactly one} unstable perturbation.

\item Static rotating solutions: The kink mean velocity is always subluminal. Thus, these solutions are \emph{always stable}.

\item Translationally invariant oscillatory solutions: In this case there are superluminal kinks with velocities ranging from $1$ to ${\bar v}_0 \left( \tilde{a}_{\max} \right)$. There are exactly two distinct kinks with the same superluminal velocity. Thus, these solutions are \emph{unstable, as long as $\beta \geq 1 / {\bar v}_0 \left( \tilde{a}_{\max} \right)$}. Whenever, they are unstable they have \emph{exactly two} unstable perturbations (except for the saturating case $\beta = 1 / {\bar v}_0 \left( \tilde{a}_{\max} \right)$, when there is only one). The two distinct modes have in general different Lyapunov exponents.

\item Translationally invariant rotating solutions: In this case there are superluminal kinks with velocities ranging from $1$ to ${\bar v}_0 \left( \omega_1 \right)$. There is only one kink for each velocity.  Thus, these solutions are \emph{unstable, as long as $\beta \geq 1 / {\bar v}_0 \left( \omega_1 \right)$}. When, they are unstable they have \emph{exactly one} unstable perturbation.
\end{enumerate}

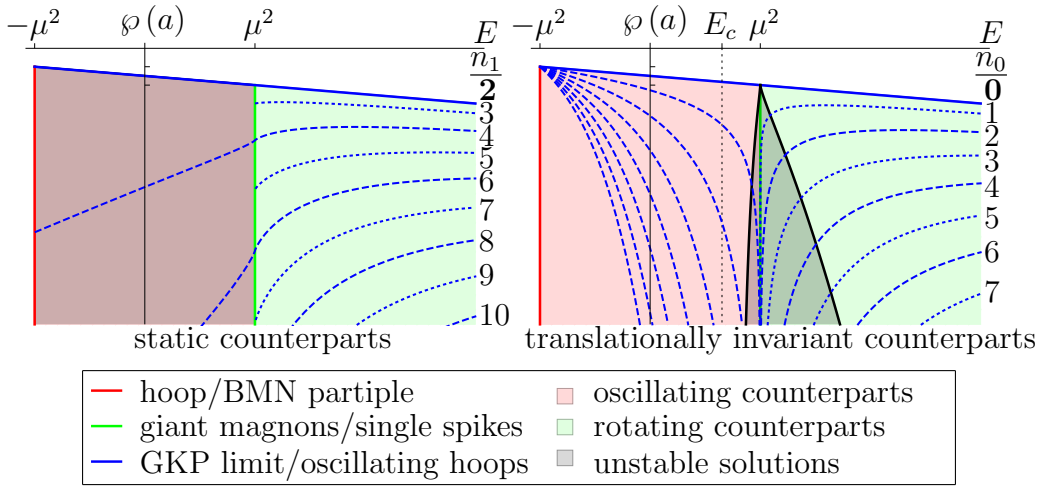
\begin{figure}[ht]
\vspace{10pt}
\begin{center}
\begin{picture}(88,42.5)
\put(1,14){\includegraphics[width = 0.4\textwidth]{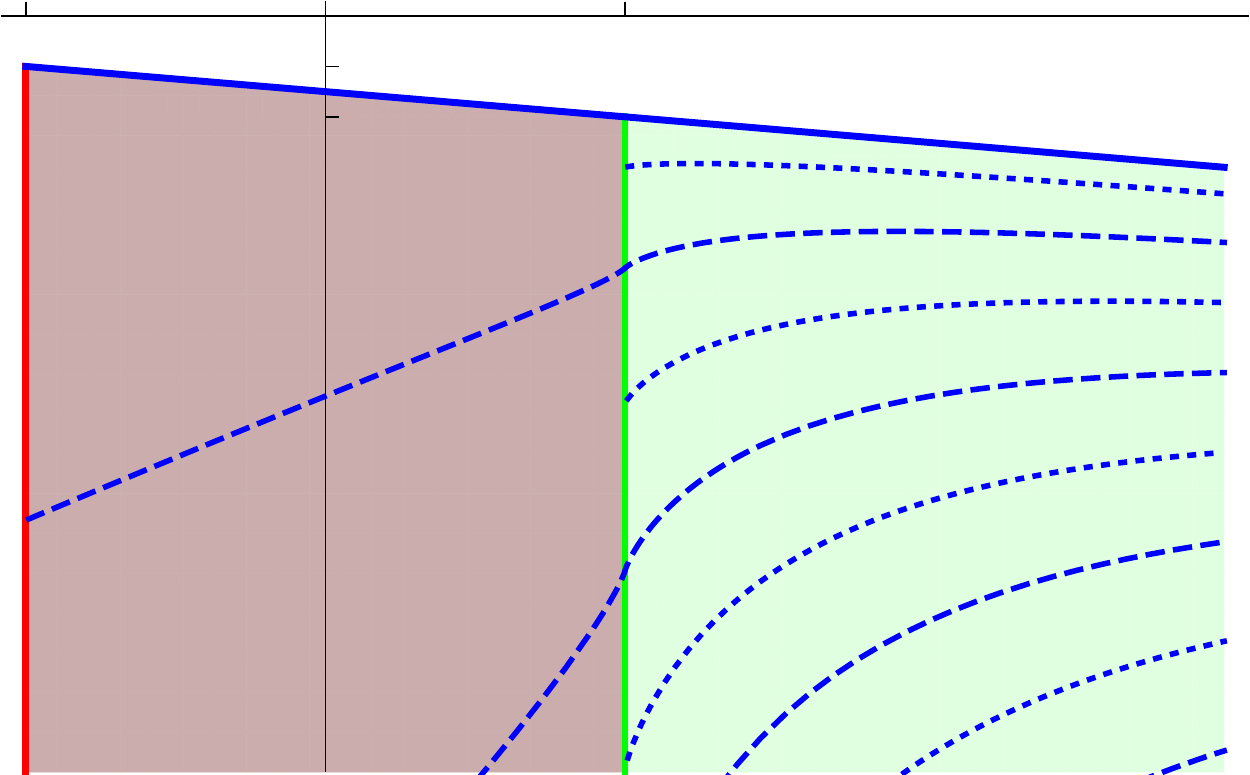}}
\put(45,14){\includegraphics[width = 0.4\textwidth]{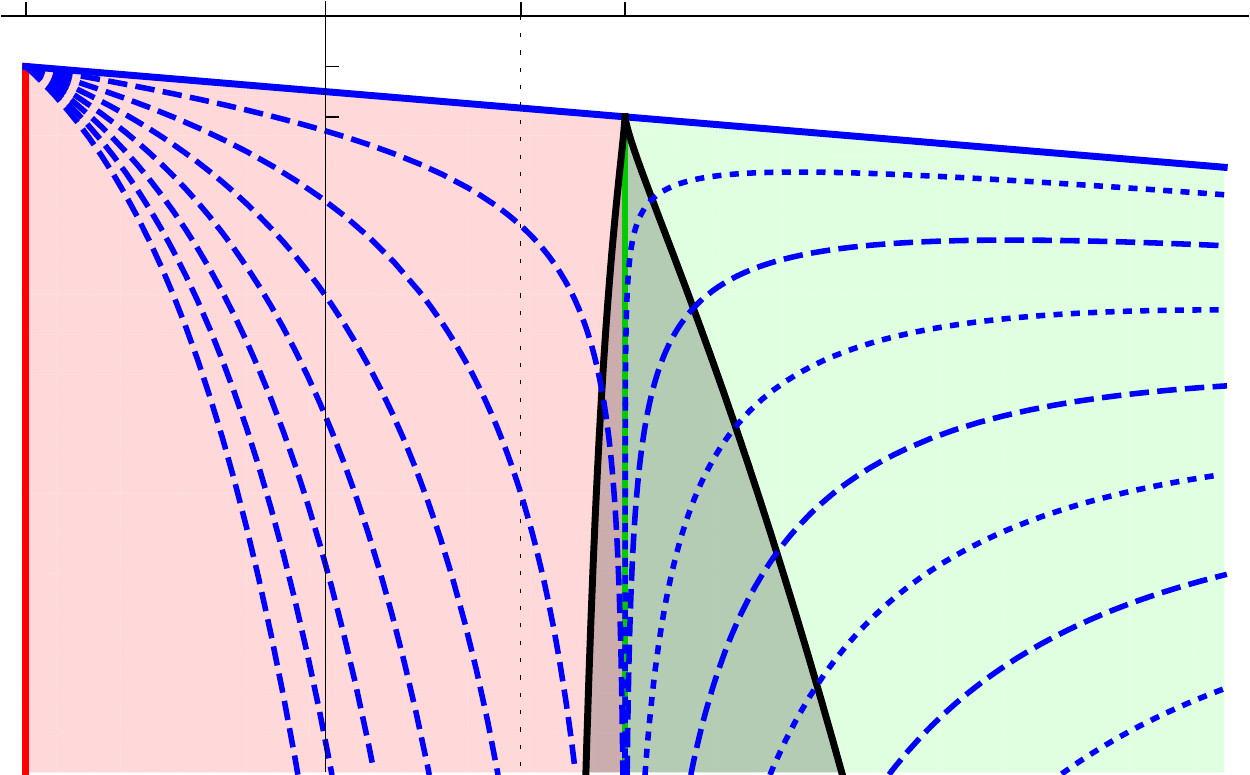}}
\put(10.5,12){static counterparts}
\put(44.5,12){translationally invariant counterparts}
\put(40,38.75){$E$}
\put(-0.75,39.5){$-\mu^2$}
\put(20,39.5){$\mu^2$}
\put(9.25,40){$\wp \left( a \right)$}
\put(39.75,36.75){$\underline{n_1}$}
\put(40.5,33.75){\bf{2}}
\put(40.5,31.75){$3$}
\put(40.5,29.75){$4$}
\put(40.5,27.75){$5$}
\put(40.5,25.75){$6$}
\put(40.5,23.25){$7$}
\put(40.5,20.5){$8$}
\put(40.5,17.5){$9$}
\put(40.5,14){$10$}
\put(84,38.75){$E$}
\put(43.25,39.5){$-\mu^2$}
\put(60,39.5){$E_c$}
\put(64,39.5){$\mu^2$}
\put(53,40){$\wp \left( a \right)$}
\put(83.75,36.75){$\underline{n_0}$}
\put(84.5,33.75){\bf{0}}
\put(84.5,31.75){$1$}
\put(84.5,29.75){$2$}
\put(84.5,27.5){$3$}
\put(84.5,25.25){$4$}
\put(84.5,22.5){$5$}
\put(84.5,19.5){$6$}
\put(84.5,16){$7$}
\put(6,0){\includegraphics[height = 0.1\textwidth]{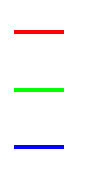}}
\put(11,1){GKP limit/oscillating hoops}
\put(11,4.25){giant magnons/single spikes}
\put(11,7.25){hoop/BMN partiple}
\put(46.5,0){\includegraphics[height = 0.095\textwidth]{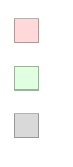}}
\put(50.5,1){unstable solutions}
\put(50.5,4.25){rotating counterparts}
\put(50.5,7.25){oscillating counterparts}
\put(6,0){\line(0,1){10}}
\put(6,0){\line(1,0){76}}
\put(82,0){\line(0,1){10}}
\put(6,10){\line(1,0){76}}
\end{picture}
\end{center}
\vspace{-10pt}
\caption{The set of unstable elliptic string solutions in the moduli space. The dashed and dotted blue lines depict the closed elliptic strings. They are indexed by the number of the real periods of the Weierstrass elliptic function that covers them.}
\vspace{5pt}
\label{fig:elliptic_string_moduli_instability}
\end{figure}

Figure \ref{fig:elliptic_string_moduli_instability} depicts the above in the moduli space of elliptic string solutions, as parametrized by the moduli $E$ and $a$.
In figure \ref{fig:elliptic_string_moduli_instability}, there are two black curves that separate the stable from the unstable solutions, in the case of translationally invariant elliptic strings. In the region $E<\mu^2$, the curve is $\beta = 1 / {\bar v}_0 \left( \tilde{a}_{\max} \right)$, which has the line $E=E_c$ as an asymptote. The strings on this curve are unstable, but with only one unstable mode. In the region $E>\mu^2$, the curve is $\beta = 1 / {\bar v}_0 \left( \omega_1 \right)$.

\section{Discussion}
\label{sec:discussion}

Recently new classical string solutions on $\mathbb{R} \times \mathrm{S}^2$ were constructed by applying the dressing method to elliptic strings \cite{Katsinis:2018ewd}. Application of the dressing method to the NLSM, is equivalent to applying \Backlund transformations on the Pohlmeyer reduced theory, namely the sine-Gordon equation in our case. In \cite{salient} it was pointed out that a particular class of the dressed elliptic strings is associated with instabilities of their elliptic seeds. In this work, we study the stability of elliptic strings in $\mathbb{R} \times \mathrm{S}^2$ by introducing linear perturbations at their Pohlmeyer counterparts. Our analysis indicates that the study of linear perturbations leads to the same conclusions as the ones obtained by the application of the dressing method.

This conclusion should not be surprising. A single \Backlund transformation adds a degenerate genus to the solution. The degenerate genus corresponds to a divergent period. Whenever one is able to align this divergent period to the temporal direction one obtains a string solution, which tends to an elliptic string in the asymptotic past and future. As a result, the existence of this kind of solutions reveals the instability of the particular seed. This reasoning can be turned on its head. If we assume the existence of an unstable seed solution, it is then the case that the solution that realizes the instability asymptotically tends to it. This implies that it should have the same genus as the seed, plus a degenerate one, whose infinite period is aligned with the time direction. It is thus natural to expect that such a solution should be accessible via a single \Backlund transformation or equivalently via the dressing method with the simplest dressing factor.

Furthermore, one should note that the dressing method not only allows the identification of the unstable strings, but it also provides the exact solution that realizes this instability. In more general setups, this is important, since one obtains the fate of the perturbations at full non-linear level. For example, in the case of the existence of metastable configurations one could probe the global stability properties, which are not accessible via the study of stability properties under small perturbations.

A key aspect of our analysis is the importance of boundary conditions. Since one should consider closed strings, the perturbations most obey appropriate periodicity conditions. The general treatment of the stability of solutions of the sine-Gordon equation \cite{SGstability} is not appropriate when one studies strings with specific topological characteristics, such as closed strings. It is interesting that the unstable closed string solutions have a very limited number of unstable modes, in our case one or two. The existence of two modes is interesting and could imply the existence of a multitude of configurations, which are either stable under small perturbations or saddle points, and act as attractors leading the system away from the unstable configuration.

The very limited number of unstable perturbations is also a characteristic of the helicoid and catenoid minimal surfaces in the hyperboloid $\mathrm{H}^3$ \cite{Wang_catenoid,Wang_helicoid}. Since these minimal surfaces can also be studied with the help of their Pohlmeyer counterparts \cite{Pastras:2016vqu}, a similar twofold (linear and through the dressing method) analysis would be interesting.

Our approach could be implemented to the study of the stability of strings which propagate in diverse background symmetric spacetimes as well (such as dS or AdS), and obviously in higher dimensions. One need not constrain the focus on string solutions of Pohlmeyer reducible systems. Whenever the dressing method is applicable, the stability of seeds can be studied in the same fashion.

\subsection*{Acknowledgements}

%

The authors would like to thank M. Axenides and E. Floratos for useful discussions.

\end{document}